# Real-space observation of charge ordering in epitaxial La$_{2-x}$Sr$_x$CuO$_4$ films

Yang Wang[1], Yong Zhong[1], Zhiling Luo[1], Menghan Liao[1], Ruifeng Wang[1], Ziyuan Dou[1], Qinghua Zhang[2], Ding Zhang[1,3], Lin Gu[2,3], Can-Li Song[1,3 †], Xu-Cun Ma[1,3 †], Qi-Kun Xue[1,3,4 †]

[1]State Key Laboratory of Low-Dimensional Quantum Physics, Department of Physics, Tsinghua University, Beijing 100084, China

[2]Institute of Physics, Chinese Academy of Sciences, Beijing 100190, China

[3]Collaborative Innovation Center of Quantum Matter, Beijing 100084, China

[4]Beijing Academy of Quantum Information Sciences, Beijing 100193, China

*To whom correspondence should be addressed to C.L.S. (clsong07@mail.tsinghua.edu.cn), X.C.M. (xucunma@mail.tsinghua.edu.cn) or Q.K.X. (qkxue@mail.tsinghua.edu.cn).

**The cuprate superconductors exhibit ubiquitous instabilities toward charge-ordered states. These unusual electronic states break the spatial symmetries of the host crystal, and have been widely appreciated as essential ingredients for constructing a theory for high-temperature superconductivity in cuprates. Here we report real-space imaging of the doping-dependent charge orders in the epitaxial thin films of a canonical cuprate compound La$_{2-x}$Sr$_x$CuO$_4$ using scanning tunneling microscopy. As the films are moderately doped, we observe a crossover from incommensurate to commensurate (4a$_0$, where a$_0$ is the Cu-O-Cu distance) stripes. Furthermore, at lower and higher doping levels, the charge orders occur in the form of distorted Wigner crystal and grid phase of crossed vertical and horizontal stripes. We discuss how the charge orders are stabilized, and their interplay with superconductivity.**

Electrons in strongly correlated materials are often susceptible to segregating into complexly ordered structures[1-12]. In particular, charge-ordered states may coexist and even compete with high temperature ($T_c$) superconductivity in moderately doped cuprates[2,6,8,9], prompting numerous investigations into their correlations with superconductivity and the mysterious "pseudogap" phenomenon[4,7]. Recently, the charge ordering was more universally confirmed to extend over a wider doping range in the cuprate phase diagram by resonant X-ray scattering[13,14], but the ultimate microscopic mechanism, its doping dependence, the exact modulated pattern (e.g., mixture of uniaxial stripes or biaxial checkerboards) and its commensurability are all as yet unclear[2-8], due in part to the twinning and phase separation in various cuprate compounds. The opportunity to probe charge ordering locally has propelled a plethora of early scanning tunneling microscopy (STM) measurements, which provide mounting evidence for short-ranged checkboard-type structure on the cleavable $Ca_{2-x}Na_xCuO_2Cl_2$ and Bi-based cuprates[2,3,15]. Nevertheless, such pattern has been suggested to be describable equivalently by nanoscale patches of uniaxial stripes as the disorder scattering is considered[16]. It would represent a significant advance to clarify the modulated charge structure of cuprates in real space.

$La_{2-x}Sr_xCuO_4$ (LSCO), the canonical cuprate superconductor with a single $CuO_2$ layer (Fig. 1a), serves as a prototype of charge ordered system. A consensus has been reached on the hole carrier configuration as nearly commensurate $4a_0$-period stripes ($a_0 \sim 0.378$ nm, the compound's in-plane lattice constant), which become prominent and overwhelm the superconductivity around $x = 1/8$[9,17]. Yet, incommensurate stripes were evident for the lower doping level[17]. Despite relentless efforts, a real-space visualization of charge ordering, indispensable for identifying its microscopic nature, is lacking, since the LSCO single crystal never exhibited good cleavage as the Bi-based cuprates. The existing STM experiments present no atomic-resolved topography on LSCO[18]. Here we combine STM with state-of-the-art ozone-assisted molecular beam epitaxy (O-MBE) to step over this long-standing obstacle by preparing high-quality LSCO films over a wide doping range (see Methods). We further image, at atomic resolution, four distinct charge-ordered structures that depend on the chemical doping. Our findings shed light into the microscopic mechanism of charge orders, as well as their interplay with the superconductivity in cuprates.

Figure 1(b) represents a typical large-scale STM topographic image of ~ 10 unit cell (UC) thick LSCO films prepared on the $SrTiO_3(001)$ substrate, displaying atomically flat morphology. Irrespective of the $Sr^{2+}$ concentration x, the terraces are uniquely separated by a half unit-cell step height of about 0.66 nm (Fig. 1c), as is expected for c(001)-orientated LSCO thin films[18,19]. This has been subsequently confirmed by our X-ray diffraction measurement in Fig. 1(d). The high crystalline quality of LSCO films is justified by the sharp Bragg peaks, from which the average c-axis lattice parameter of ~ 1.314 nm is extracted. Given the rigidity of $CuO_6$ octahedron, we can reasonably attribute the exposed surface as LaO termination (see the inset of Fig. 1c)[18]. Further structural analysis by high-resolution transmission electron microscopy (STEM) reveals a buffer layer of another perovskite oxide $LaCuO_3$ sandwiched between the epitaxial films and $SrTiO_3$ substrate[19], as evident from Fig. 1e. This might be more probably driven by the large lattice mismatch of 3% in $LSCO/SrTiO_3$(001) heteroepitaxy. The in-plane lattice parameter (0.382 nm) of the $LaCuO_3$ buffer layer lies between those of $SrTiO_3$ (0.3905 nm) and LSCO (0.378 nm), and its insertion would effectively release the epitaxial stress that optimizes the epitaxial growth of LSCO films on $SrTiO_3$.

Notably, the crystalline quality of LSCO films changes little with x in spite of the fact that the partial substitution of $La^{3+}$ by $Sr^{2+}$ species introduces hole (p)-type dopants into the films. In order to explore the charge ordering over a wider doping range, we have also prepared slightly La-rich $La_2CuO_4$ (LCO) films (x = 0, La:Cu ~ 2.1), in which the excess La cations compensate the oxygen anions inevitably occurring at interstitial sites, pushing the samples close to the "parent" state. Figure 2 shows the atomically-resolved STM images and electronic structures of LSCO films as a function of carrier concentration p. It is worth noting that the absolute doping p is currently inaccessible, since at a growth temperature of 700°C the Sr atoms are very volatile and the real flux rate exhibits great deviation from the calibrated one at room temperature by the quartz crystal microbalance. At the extremely low doping level (Fig. 2a), the $La_{2+\delta}CuO_4$ films ($\delta$ ~ 0.1) are totally insulating (the top panel of Fig. 2f) and display a slightly distorted hexagonal superstructure, marked by the white rhombus. We measure its period to be ~ 1.56 nm = $4a_0$. Increasing the doping level by the substitution of a small amount of $La^{3+}$ for $Sr^{2+}$ converts the distorted hexagonal pattern to stripe-like superstructures (Fig.

2b), which run exclusively along the crystallographic axes *a* or *b*. These stripes are not uniformly spaced and mostly incommensurate with the crystal lattice $a_0$ (Supplementary Fig. 1), which we dub as incommensurate stripes. Further doping of $Sr^{2+}$ leads to commensurate stripes with a modulation period of $4a_0$. This can be inferred from the overlaid atomic lattice in Fig. 2c. Finally, at the higher doping levels, following a surface avoid of large superstructure (Fig. 2d), the LSCO surface becomes characteristic of crossed vertical and horizontal stripes that we name as grid phase with an averaged spacing of ~ 4.01 nm (Fig. 2e and Supplementary Fig. 2). Note that the adjacent bright spots in both Figs. 2d and 2e are spaced 0.53 nm. This matches with the lattice parameter of a $\sqrt{2} \times \sqrt{2}$ reconstructed surface (marked by yellow squares), which we suggest is driven by the polar compensation of $LaO^{1+}$-$CuO_2^{2-}$-$LaO^{1+}$ stacking sequence. Remarkably, the $\sqrt{2} \times \sqrt{2}$ reconstruction seems generic for all the observed surfaces other than the striped ones, in which the $a_0$-spaced La atoms are observable along the stripes (Figs. 2b and 2c). In Figs. 2f-2j, we plot the differential conductance *dI/dV* spectra on various surfaces. It is immediately found that the spectral gaps diminish with increasing *p*, a hallmark of dopant-triggered Mott-insulator-metal transition. It should be cautioned that the measured gap might be overestimated by the combined effects of tip-induced band bending[20] and $LaCuO_3$ buffer layer. This appears consistent with the gap remnants in LSCO films with sufficiently high doping levels (Figs. 2c-2e).

To understand the varying patterns in Figs. 2a-e, we first focus on the $4a_0$-period commensurate stripe (Fig. 3a). The stripes are characteristic of bidirectional orientation, which run along either *a* or *b* axis and extend over a few tens of nanometers. Figure 3b plots the conductance *dI/dV* spectra both on and off the stripes, revealing a sharp distinction. The different shape of *dI/dV* on and off the stirpes presents the first evidence that the formation of stripes may be electronically driven. This is corroborated by the energy dependent *dI/dV* conductance maps (Fig. 3c and Supplemental Fig. 3), in which a reversal of spatially-resolved *dI/dV* amplitude becomes obvious as a function of energy. Besides we have measured the variation of tunneling current *I* with varying tip-sample distance *z* (Fig. 3d), based on which the local tunneling barrier $\phi =$ $0.952 \times (\frac{dlnI}{dz})^2$ is calculated. Again, the same $4a_0$-periodic spatial modulation in $\phi$ is revealed from Fig. 3e. As compared to the inter-stripe regions, the local tunneling barrier $\phi$, or roughly the local work function, is apparently larger on the stripes. These findings hint at that the $4a_0$-period stripes observed here most probably correspond to the charge order previously studied in LSCO[1,9,17]. If so, our current study represents the real-space observation of it in the canonical cuprate compound LSCO.

Beyond the stripe phases, we have further revealed that the other superstructures in Fig. 2, to wit the distorted hexagonal pattern and grid phase, are of electronic origin as well (Fig. 4 and Supplemental Fig. 4). As displayed in Figs. 4a-c, for example, the grid phase changes with the sample voltage and is characteristic of an intriguing corrugation reversal in the STM topographic images. Meanwhile, half of La atoms become invisible as the energy goes higher, leading to a crossover of the surface pattern from $\sqrt{2} \times \sqrt{2}$ to $2 \times 2$. A similar tunneling barrier measurement exhibits a higher $\phi$ (~ 4.7 eV) of the grid edges than that (~ 4.1 eV) of hollow regions as shown in Fig. 4d. All the results support the electronic origin of the observed superstructures, which we interpret as various forms of charge orders. Indeed, the corrugation amplitudes of these superstructures have been measured to be as high as 0.2 nm in Fig. 2 at a setpoint voltage of -5.0 V, which is rather large and cannot be interpreted as structural distortion. Otherwise, our TEM images should have already resolved the modulations.

Our real-space imaging of epitaxial LSCO films provides an atomic-scale basis for the description of charge orders, which constitutes the major finding in this study. More significantly, the characterization of different charge-ordered patterns at various doping levels would certainly shed light on their microscopic mechanism. We believe it unlikely that the observed patterns are related to the lattice mismatch between the LSCO overlayer and substrate, as they behave quite consistently among various samples (> 130) with varied thicknesses (i.e. Supplemental Fig. 5). Each type of charge order, which has been observed in at least 10 samples, seems to be inherent for LSCO films with a certain range of composition (i.e. Sr/La ratio). In diluted antiferromagnet of LCO (Fig. 2a), the averaged separation between the charge carriers appears relatively huge and the mutual long-range Coulomb repulsion might play a predominant role in charge configuration. Intuitively, an isotropic (Wigner crystalline order) organization of charges is energetically desired. Such a hexagonal charge order, however, may be somewhat frustrated by the underlying square lattice[21], leading to the Wigner crystal-like distorted hexagonal pattern in Fig. 2a. Actually, the charge pattern is found to be commensurate with the LSCO lattice along one of the crystalline axes on the *ab*-plane.

Furthermore, in LSCO compounds with strong antiferromagnetic correlations, the doped holes can distort the surrounding spin configuration, giving rise to dipole-dipole attraction interactions between them. They compete with the long-range Coulomb interactions and allow the generation of a variety of superstructures at the increased chemical doping[22-25], including the striped (Figs. 2b and 2c) and grid (Fig. 2e) phases observed above. In theory, multiple models have been employed to explain the striped phases, irrespective of whether or not the long-range coulomb interactions are treated on an equal footing[25,26]. However, we stress that, at least, the crossover from incommensurate (Fig. 2b) to commensurate (Fig. 2c) stripes, consistent with the previous doping-dependent experiments[17], is probably triggered by the long-range Coulomb repulsion. As played in the Wigner crystal phase, it not only drives the charge inhomogeneity, but also configures the striped charge distribution into a well-ordered manner (Fig. 2c). It is worth noting that the stripes invariably run along the crystal axes and exist over a wide doping range, which counters the scenario of diagonal stripes in underdoped LSCO cuprates[22,27]. This discrepancy may origin from the sample diversity. Nevertheless, our finding implies that the vertical stripes are more favorable, in agreement with a recent numerical simulation based on the Hubbard model[26]. On further hole doping, the antiferromagnetic correlations in LSCO become weaker and the striped phases might become unstable. As a consequence, a new grid phase of crossed vertical and horizontal stripes emerges in Fig. 2e[23], although a further theoretical simulation is needed to fully understand how it develops from the subtle competition between the long-range Coulomb repulsion and dipole-dipole attraction interactions.

Next we would address the relationship between the observed charge orders and superconductivity. Our transport measurements revealed that the LSCO films studied above exhibit no signature of superconductivity, whereas the LSCO films prepared on another substrate LaSrAlO$_3$ (LSAO, $a_0 \sim 0.377$ nm) via the identical growth procedure are superconducting at the appropriate doping of Sr$^{2+}$ species (Supplemental Fig. 6). As thus, the occurrence of strong charge orders in LSCO/SrTiO$_3$ films supports a competing scenario of the charge orders with superconductivity[9].

Distinct from previous STM measurements mostly devoted to low-energy states of cuprates[2,3,7,15], we collect the extremely high-energy ones of LSCO by applying large voltage to the samples. This enables us to not only overcome the difficulty in studying the insulating states of cuprates, but also access directly the Mott-Hubbard bands from which the high-$T_c$ superconductivity develops[28]. In sense, our findings hint at that the charge ordering might be inherent to the Mott states, although its behavior varies with doping (Fig. 2). As compared to the dopant-induced low-energy states[3,7,28], the intrinsic Mott states are less prone to spatial inhomogeneity, in accord with the observed strong, relatively homogeneous charge orders in Fig. 2.

Our STM study resolves several issues regarding the charge ordering of doped Mott insulators, for example, how it behaves itself on the atomic scale and how it evolves with doping. The results reported here reveal that the charge ordering relies substantially on doping level and behaves quite complex. Moreover, the absence of $\sqrt{2} \times \sqrt{2}$ surface reconstruction in the striped phases and the transition of surface structure from $\sqrt{2} \times \sqrt{2}$ to $2 \times 2$ in the grid phase suggest an intricate interplay of charge ordering with the lattice degrees of freedom. It raises a question whether the observed charge orders represent the bulk property or are stabilized only near the surface[9,29], which merits further theoretical and experimental exploration. In any case, the observations of various charge orders in LSCO films are consistent with most theoretical proposals that the doped Mott insulators are preferably charge-ordered[23-26]. Our real-space characterization of charge orders provides the atomic-scale information for uncovering its microscopic mechanism and imposes strong constraints on any theoretical models of charge ordering in cuprates.

**METHODS**

**Sample growth.** High-quality $La_{2-x}Sr_xCuO_4$ films were prepared in an ozone assisted molecular beam epitaxy (the base pressure is $1.0 \times 10^{-10}$ torr) chamber, equipped with a quartz crystal microbalance (QCM, Inficon SQM160H) for flux calibration. The ozone flux was injected from our home-built gas delivery system into the O-MBE chamber via a nozzle 50 mm away from the substrates. Atomically flat $SrTiO_3$(001) substrates were cleaned by being heated to 1200°C under ultrahigh vacuum (UHV) for 10 minutes, rendering a $TiO_2$ termination layer, while the LSAO(001) substrates were *ex-situ* annealed in a tube furnace. The LSAO substrates were then transferred into the O-MBE chamber and degassed at 400°C for 2 hours in UHV to wipe off the surface adsorbates. The deposition processes were achieved by co-evaporation of high-purity metal sources (La, Sr and Cu) from standard Knudsen cells under an ozone flux beam of $\sim 1.5 \times 10^{-5}$ Torr, on Nb-doped $SrTiO_3$(001) or insulating LSAO(001) substrates kept at 700°C. Flux of the metal sources was precisely calibrated in sequence prior to every film epitaxy, ensuring a growth rate of 6 min per unit cell. After growth, the samples were first cooled down to 300°C under the same $O_3$ atmosphere to protect the films from decomposition, and then to room temperature in UHV to guarantee the surface cleanliness.

***In-situ* STM measurements.** Our STM measurements were carried out in a Unisoku USM 1300S $^3$He system at a fixed temperature of 4.2 K. Polycrystalline PtIr tips were pre-cleaned via e-beam heating in UHV, calibrated on MBE-prepared Ag/Si(111) films, and then used throughout the experiments. All the STM topographies were acquired in a constant current mode with the voltage applied on the sample. The *dI/dV* spectra and maps were measured by using a standard lock-in technique with a small bias modulation of 10 meV at 919 Hz. The *I-z* spectroscopy was obtained by recording the tunneling current with varying tip-sample separation *z*, with the feedback loop switched off.

**Ex-situ XRD, STEM and transport measurements.** After STM characterizations, we transferred the samples out of the UHV chamber for ex-situ XRD, STEM and transport measurements. The XRD measurements of epitaxial $La_{2-x}Sr_xCuO_4$ films were carried with a high-resolution diffractometer (Rigaku, Smartlab) using the monochromatic Cu K$_{\alpha 1}$ radiation ($\lambda$=1.54178Å). After this, we investigated the crystalline structures with a high resolution STEM (JEM-ARM200F and G2 60-300S/TEM), operated at 200-300 kV and equipped with double-aberration correctors. The available point resolution is better than 0.08 nm. The low-temperature transport properties of superconducting La$_{2-x}$Sr$_x$CuO$_4$ films on LaSrAlO$_4$(001) were characterized in a closed-cycle system (Oxford Instruments TelatronPT), equipped with a He-3 insert (a base temperature of 0.25 K). The temperature sensor was placed directly below the sample stage and positioned in a configuration with the minimal magnetoresistance. Freshly-cut indium cubes were cold pressed onto the sample as contacts. Standard lock-in technique was employed to measure the sample resistance in a four-terminal configuration with an excitation current of 1 µA at 13 Hz. The magnetic field was oriented parallel to the *c*-axis of the epitaxial films.

**DATA AVAILABILITY**

All data that are present here and support the conclusions of this study are available from the corresponding author on reasonable request.

**ACKNOWLEDGEMENTS**

We thank Andrey Oreshkin (MSU), Zheng-Yu Weng (IAS, Tsinghua University) and Fu-Chun Zhang (Kavli ITS, UCAS) for helpful discussions. This work was financially supported by the Ministry of Science and Technology of China (2015CB921001, 2017YFA0304600, 2018YFA0305603, 2016YFA0301004), the National Natural Science Foundation of China (grant no. 11427903, 11504196, 11634007), and in part by the Beijing Advanced Innovation Center for Future Chip (ICFC).

**AUTHOR CONTRIBUTIONS**

X.C.M. and Q.K.X. conceived the experiments. Y.W., C.L.S., and X.C.M. performed the experiments. Y.Z., Z.L.L., R.F.W., and Z.Y.D. assisted the experiments. M.H.L. and D.Z. performed the transport measurement. Q.H.Z. and L.G. performed the TEM experiment. Y.W., C.L.S., and X.C.M. analyzed the data. C.L.S. and W.Y. wrote the paper with input from X.C.M. and Q.K.X.. All authors discussed the results and commented on the manuscript.

**References**

1. Tranquada, J. M., Sternlleb, B. J., Axe, J. D., Nakamura, Y. & Uchida, S. Evidence for stripe correlations of spins and holes in copper oxide superconductors. *Nature* **375**, 561–563 (1995).

2. Howald, C., Eisaki, H., Kaneko, N. & Kapitulnik, A. Coexistence of periodic modulation of quasiparticle states and superconductivity in $Bi_2Sr_2CaCu_2O_{8+\delta}$. *Proc. Natl Acad. Sci. USA* **100**, 9705–9709 (2003).

3. Hanaguri, T. *et al.* A "checkerboard" electronic crystal state in lightly hole-doped $Ca_{2-x}Na_xCuO_2Cl_2$. *Nature* **430**, 1001–1005(2004).

4. Wu T. *et al.* Magnetic-field-induced charge-stripe order in the high-temperature superconductor $YBa_2Cu_3O_y$. *Nature* **477**, 191–194 (2011).

5. Blackburn, E *et al.* X-Ray diffraction observations of a charge-density-wave order in superconducting ortho-II $YBa_2Cu_3O_{6.54}$ single crystals in zero magnetic field. *Phys. Rev. Lett.* **110**, 137004 (2013).

6. Ghiringhelli, G. *et al.* Long-range incommensurate charge fluctuations in (Y, Nd)$Ba_2Cu_3O_{6+x}$. *Science* **337**, 821–825 (2012).

7. Parker, C. V *et al.* Fluctuating stripes at the onset of the pseudogap in the high-Tc superconductor $Bi_2Sr_2CaCu_2O_{8+x}$. *Nature* **468**, 677–680 (2010).

8. Torchinsky, D. H., Mahmood, F., Bollinger, A. T., Božović, I. & Gedik, N. Fluctuating charge-density waves in a cuprate superconductor. *Nature Mater.* **12**, 387–391 (2013).

9. Croft, T. P., Lester, C., Senn, M. S., Bombardi, A. & Hayden, S. M. Charge density wave fluctuations in $La_{2-x}Sr_xCuO_4$ and their competition with superconductivity. *Phys. Rev. B* **89**, 224513 (2014).

10. Tabis, W. *et al.* Charge order and its connection with Fermi-liquid charge transport in a pristine high-$T_c$ cuprate. *Nat. Commun.* **5**, 5875 (2014).

11. Fradkin, E., Kivelson, S. A. & Tranquada, J. M. Colloquium: Theory of intertwined orders in high temperature superconductors. *Rev. Mod. Phys.* **87**, 457–482 (2015).

12. Dagotto, E. Complexity in strongly correlated electronic systems. *Science* **309**, 257–262 (2005).

13. Da Silva Neto, E. H. *et al*. Charge ordering in the electron-doped superconductor $Nd_{2-x}Ce_xCuO_4$. *Science* **347**, 282–285 (2015).

14. Peng, Y. Y. *et al*. Re-entrant charge order in overdoped $(Bi,Pb)_{2.12}Sr_{1.88}CuO_{6+\delta}$ outside the pseudogap regime. *Nature Mater.* **17**, 697–702 (2018).

15. Hoffman, J. E. et al. A four unit cell periodic pattern of quasi-particle states surrounding vortex cores in $Bi_2Sr_2CaCu_2O_{8+\delta}$. *Science* **295**, 466–469 (2002).

16. Robertson, J. A., Kivelson, S. A., Fradkin, E., Fang, A. C. & Kapitulnik, A. Distinguishing patterns of charge order: stripes or checkerboards. *Phys. Rev. B* **74**, 134507 (2006).

17. Yamada, K. *et al*., Doping dependence of the spatially modulated dynamical spin correlations and the superconducting-transition temperature in $La_{2-x}Sr_xCuO_4$. *Phys. Rev. B* **57**, 6165–6172 (1998).

18. Kato, T., Okitsu, S. & Sakata, H. Inhomogeneous electronic states of $La_{2-x}Sr_xCuO_4$ probed by scanning tunneling spectroscopy. *Phys. Rev. B* **72**, 144518 (2005).

19. Houben, L. Aberration-corrected HRTEM of defects in strained $La_2CuO_4$ thin films grown on $SrTiO_3$. *J. Mater. Sci.* **41**, 4413–4419 (2006).

20. Battisti, I. *et al*., Poor electronic screening in lightly doped Mott insulators observed with scanning tunneling microscopy. *Phys. Rev. B* **95**, 235141 (2017).

21. Rademaker, L., Pramudya, Y., Zaanen, J. & Dobrosavljević, V. Influence of long-range interactions on charge ordering phenomena on a square lattice. *Phys. Rev. E* **88**, 032121 (2013).

22. Seul, M. & Andelman, D. Domain shapes and patterns: the phenomenology of modulated phases. *Science* **267**, 476-483 (1995).

23. Stojković, B. P. *et al*., Charge ordering and long-range interactions in layered transition metal oxides: A quasiclassical continuum study. *Phys. Rev. B* **62**, 4353–4369 (2000).

24. Kou, S. P. & Weng, Z. Y. Holes as dipoles in a doped antiferromagnet and stripe instabilities. *Phys. Rev. B* **67**, 115103 (2003).

25. Seibold, G., Castellani, C., Castro, C. D. & Grilli, M. Striped phases in the two-dimensional Hubbard model with long-range Coulomb interaction. *Phys. Rev. B*

**58**, 13506–13509 (1998).

26. Zheng, B. X. *et al.*, Stripe order in the underdoped region of the two-dimensional Hubbard model. *Science* **358**, 1155–1160 (2008).

27. Matsuda, M. *et al.*, Static and dynamic spin correlations in the spin-glass phase of slightly doped $La_{2-x}Sr_xCuO_4$. *Phys. Rev. B* **62**, 9148–9154 (2000).

28. Ye, C. *et al.*, Visualizing the atomic-scale electronic structure of the $Ca_2CuO_2Cl_2$ Mott insulator. *Nat. Commun.* **4**, 1365 (2013).

29. Wu, H. H. *et al.*, Charge stripe order near the surface of 12-percent doped $La_{2-x}Sr_xCuO_4$. *Nat. Commun.* **3**, 1023 (2012).

**Figure Legends**

**Figure 1 | Characterization of LSCO films on SrTiO$_3$. a**, Crystal structure of LSCO. The lattice is defined with respect to the conventional tetragonal unit cell (space group I4/mmm, lattice parameters $a \sim 0.378$ nm, $c \sim 1.32$ nm). **b**, STM topography ($V$ = -5.0 V, $I$ = 10 pA, 500 nm × 500 nm) of ~ 10 UC thick LSCO films. **c**, Line-cut profile along the white curve in **a**, revealing a unique step height of ~ 6.57 Å. Inset shows the LaO-terminated surface. **d**, X-ray diffraction (XRD) pattern of the representative LSCO films. The sharp Bragg peaks characteristic of both LSCO films and SrTiO$_3$ substrate are indicated. **e**, TEM image showing the epitaxial relationship of LSCO films on SrTiO$_3$. The scale bar is 1 nm in width. For clarify, color-coded atoms matching with the experimental observations are partly superimposed on the image. A buffer layer of LaCuO$_3$ with a thickness of 4 ~ 5 UCs is clearly seen.

**Figure 2 | Doping dependence of charge ordering and electronic structures. a-e**, Various charge orders with increasing doping $p$ in LSCO. The white rhombus, with an acute interior angle of 68°, indicates the formation of distorted Wigner crystal, while the squares mark the unit cell of $\sqrt{2} \times \sqrt{2}$ surface reconstruction throughout. Colored spheres, each of which denotes an atom as specified in Fig. 1a, are overlaid on the surfaces to exemplify the 4$a_0$-period commensurate stripes in **c** and $\sqrt{2} \times \sqrt{2}$ reconstruction in **d**. Tunneling conditions: **a**, $V$ = -4.0 V, $I$ = 40 pA; **b**, $V$ = -3.6 V, $I$ = 20 pA; **c**, $V$ = -3.0 V, $I$ = 60 pA; **d**, $V$ = -2.4 V, $I$ = 15 pA; **e**, $V$ = -2.0 V, $I$ = 20 pA. The scale bar widths are 2 nm in **a**, 5 nm in **b** and 1 nm in **c-e**. **f-j**, Spatially-averaged $dI/dV$ spectra on various charge-ordered phases as indicated. The setpoints are kept to the starting voltages for every voltage ramp from the negative side, while the current $I$ is fixed at 200 pA except for the $\sqrt{2} \times \sqrt{2}$ reconstructed surface ($I$ = 100 pA).

**Figure 3 | Electronic structure of the 4$a_0$-period commensurate stripe. a**, Large-scale STM topography ($V$ = -5.0 V, $I$ = 10 pA, scale bar width 10 nm) showing bidirectional charge stripes, marked by the white dashes. **b**, Typical $dI/dV$ conductance spectra on and off the stripes. Each of them corresponds to the mean of 32 point

spectra from the same field of view. Setpoint: $V$ = -2.4 V, $I$ = 100 pA. **c**, Differential conductance maps recorded at different energies as indicated ($V$ = -2.4 V, $I$ = 10 pA, 7 nm × 7 nm). **d**, Tunneling current measured as a function of the varied tip-sample distance $\Delta z$. Here a positive $\Delta z$ means an increased tip-sample distance. As anticipated the tunneling current $I$ reduces exponentially with $\Delta z$. **e**, Map (5 nm × 7 nm) of local tunneling barrier. As compared to the inter-stripe regions, a larger work function is found on the stripes.

**Figure 4 | STM imaging of the grid phase. a-c**, Bias-dependent STM topographic images. The scale bar is 1 nm in width and tunneling conditions are set at **a**, $V$ = -2.0 V; **b**, $V$ = -0.4 V; **c**, $V$ = 1.0 V and a fixed current of $I$ = 40 pA. **d**, Local tunneling barrier measurements on the edges (red curve) and hollow regions (black curve) of grid phase, as indicated in **a**.

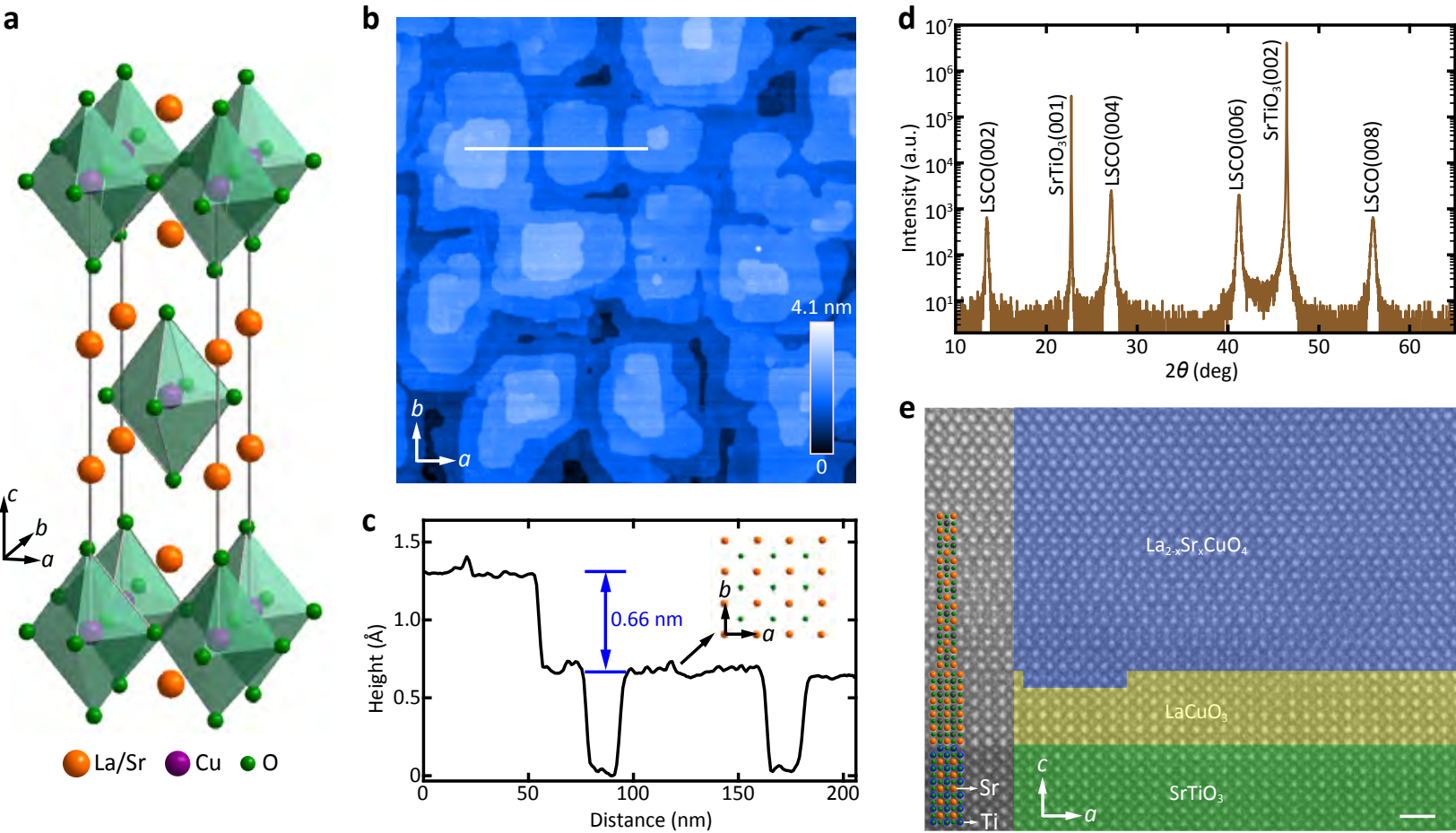

Figure 1

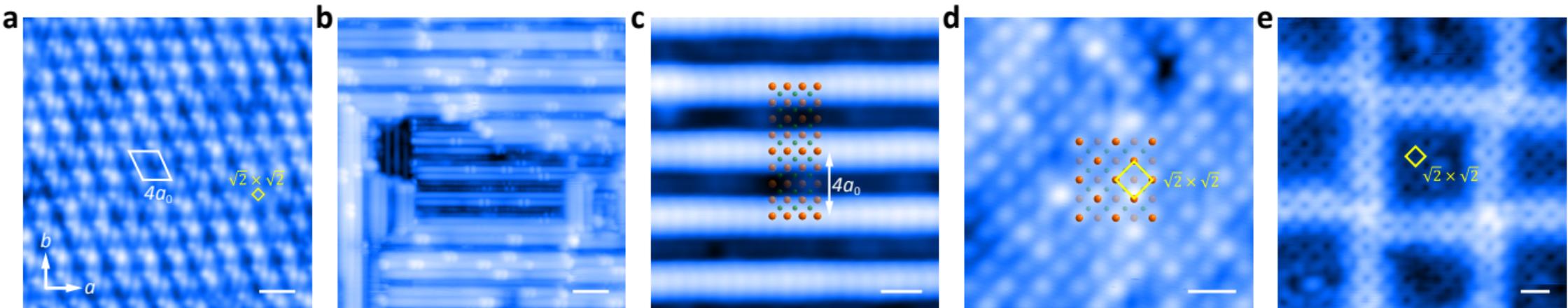
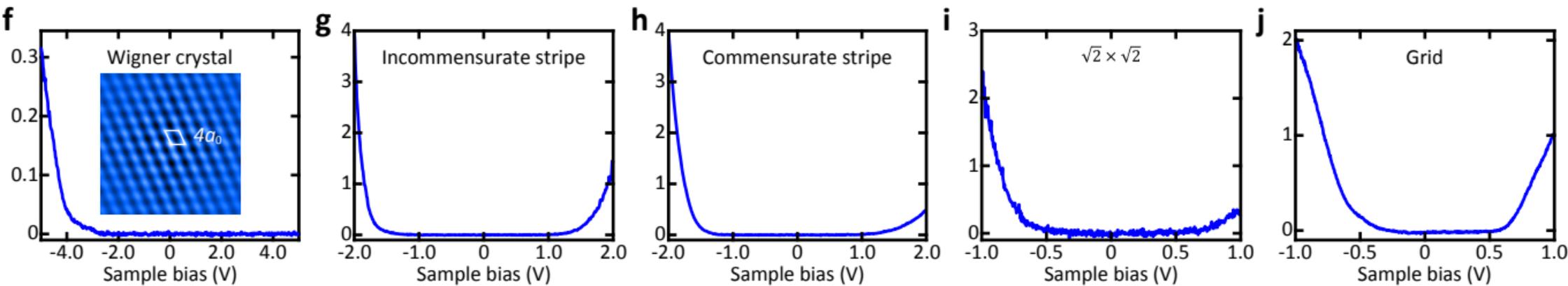

Figure 2

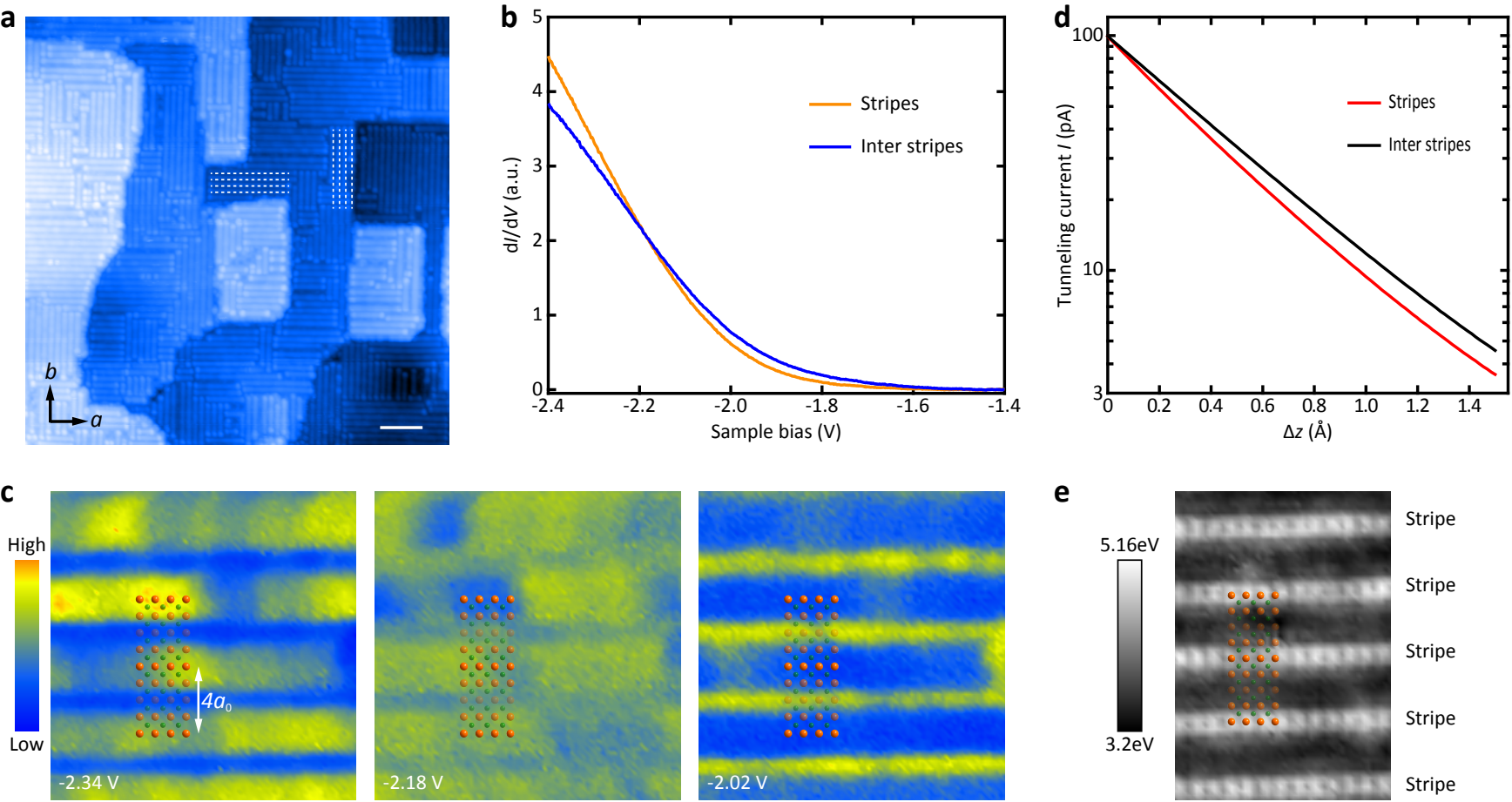

Figure 3

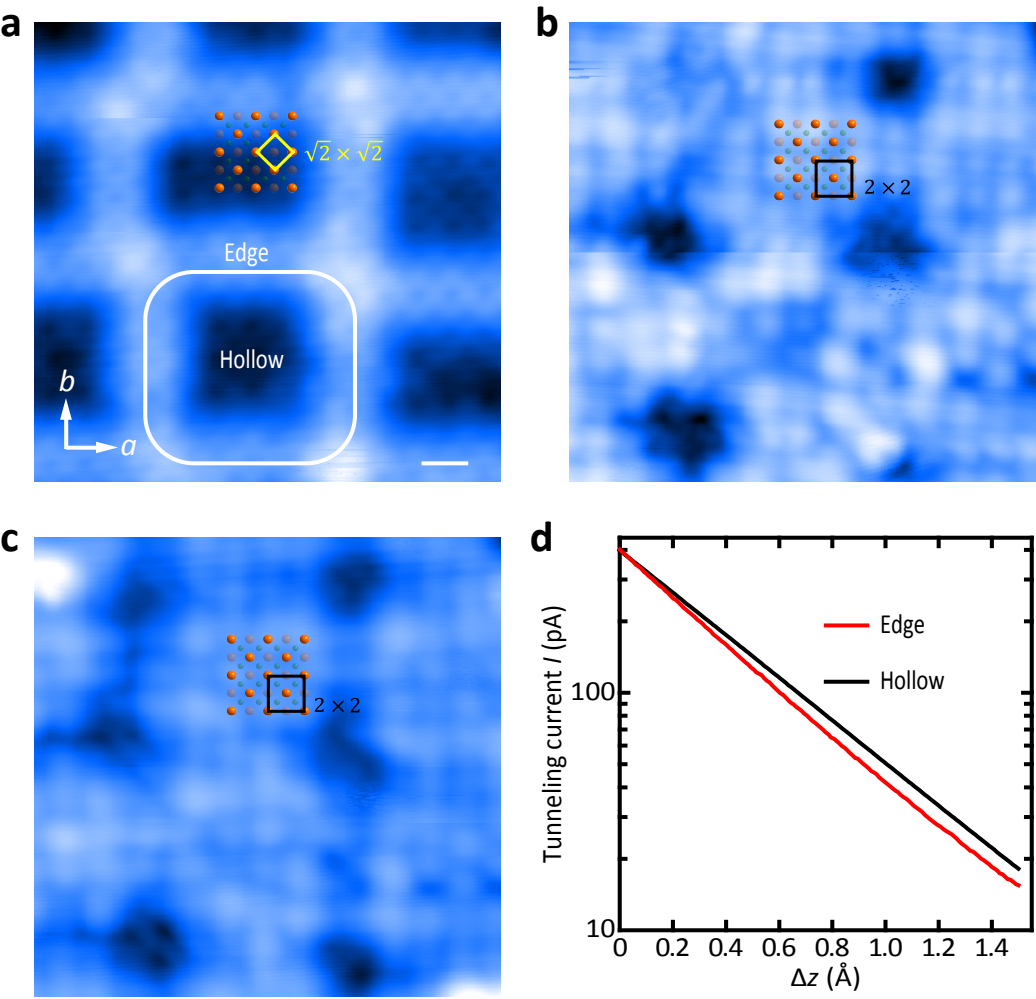

Figure 4

## Supplementary Materials for

### Real-space observation of charge ordering in epitaxial La$_{2-x}$Sr$_x$CuO$_4$ films

Yang Wang[1], Yong Zhong[1], Zhiling Luo[1], Menghan Liao[1], Ruifeng Wang[1], Ziyuan Dou[1], Qinghua Zhang[2], Ding Zhang[1,3], Lin Gu[2,3], Can-Li Song[1,3 †], Xu-Cun Ma[1,3 †], Qi-Kun Xue[1,3,4 †]

[1]State Key Laboratory of Low-Dimensional Quantum Physics, Department of Physics, Tsinghua University, Beijing 100084, China

[2]Institute of Physics, Chinese Academy of Sciences, Beijing 100190, China

[3]Collaborative Innovation Center of Quantum Matter, Beijing 100084, China

[4]Beijing Academy of Quantum Information Sciences, Beijing 100193, China

S1. Separation between neighboring stripes of the incommensurate stripe phase

S2. Spatial periodicity of the grid phase

S3. Energy dependent *dI/dV* maps for the 4$a_0$-period stripe phase

S4. Bias-dependent STM topographies of the Wigner crystal phase

S5. Incommensurate stripe phase on LSCO films with varying thicknesses

S6. Characterization of the LSCO films on LSAO substrate

## S1. Separation between neighboring stripes of the incommensurate stripe phase

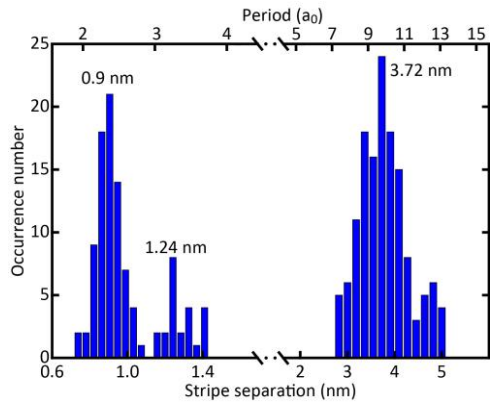

Figure S1. Statistics of the separation between adjacent stripes in the incommensurate stripe phase and the calculated occurrence number. The statistics involves twelve STM images from different regions and three samples with the similar LSCO composition. This leads to three inter-stripe spacing as marked, which are all incommensurate with the crystal lattice $a_0$.

S2. Spatial periodicity of the grid phase

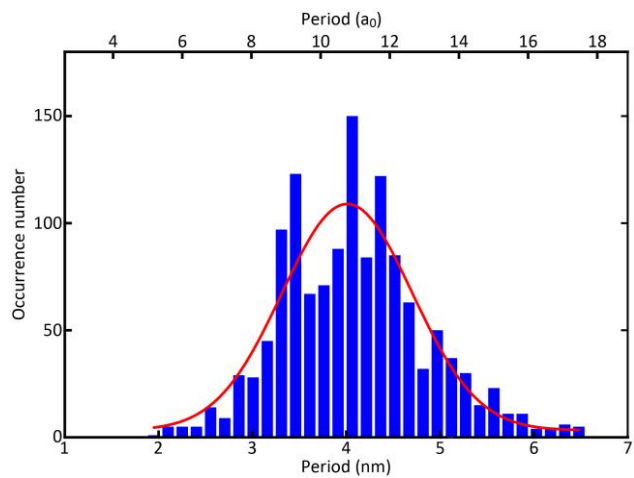

Figure S2. Histogram showing the spatial periodicity of grid phase. The periods were measured in real space, involving > 50 STM images with various Sr doping. Evidently the spatial periods of the grid phase are peaked at around 4.01 nm and conform to a Gaussian distribution (red curve).

S3. Energy dependent dI/dV maps for the 4$a_0$-period stripe phase

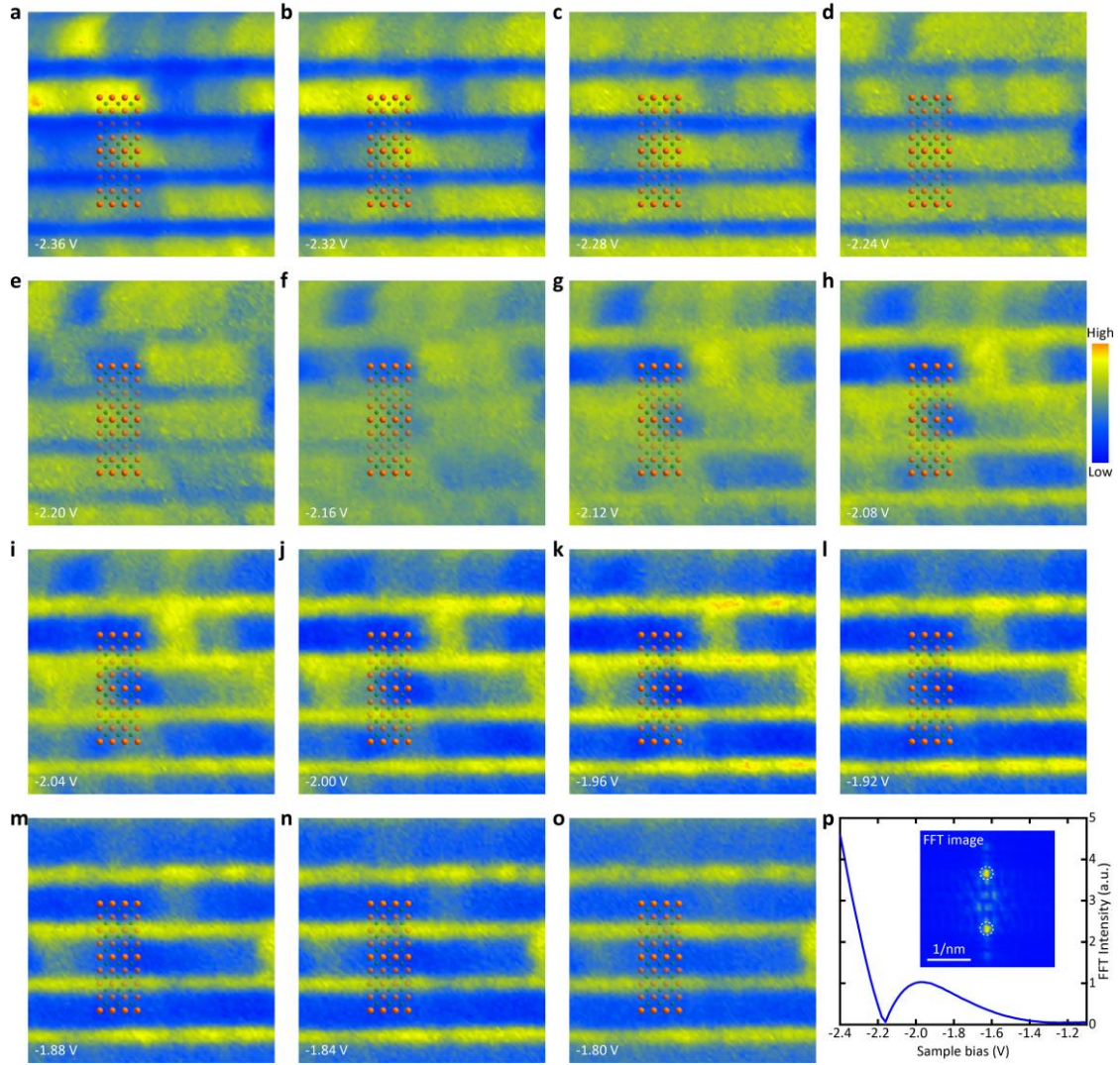

Figure S3. a-o, Differential conductance maps (7 nm × 7 nm) of the 4$a_0$-period stripes at varying energies as indicated. Setpoint: $V$ = -2.4 V, $I$ = 100 pA. p, Energy-dependent FFT intensity for the 4$a_0$-period stripes, measured from the Fourier transform images of the above dI/dV maps. Inset illustrates a typical FFT image at -2.36 eV. The bright spots circled in white originate from the real-space 4$a_0$-period stripes.

To demonstrate the electronic origin of the 4$a_0$-period stripe phase, we measured the differential conductance dI/dV maps as a function of sample voltage, ranging from -2.4 eV to -1.1 eV. As exemplified in Figs. 3c and S3a-o, these dI/dV maps change profoundly with energy and exhibit a spatial reversal of dI/dV contrast as a function of energy, a hallmark of electronic nature for the commensurate stripes. This is confirmed by the fast Fourier transform (FFT) of *dI/dV* maps, with one of them inserted in Fig. S3p. Two bright spots circled in white are observed to originate from the $4a_0$-period stripes. The intensity of FFT peaks characterizes the strength of spatially-modulated stripes. Figure S3p summarizes the energy dependence of FFT intensity for the $4a_0$-period stripes, proving a minimal value, where the spatial *dI/dV* contrast reverses.

S4. Bias-dependent STM topographies of the Wigner crystal phase

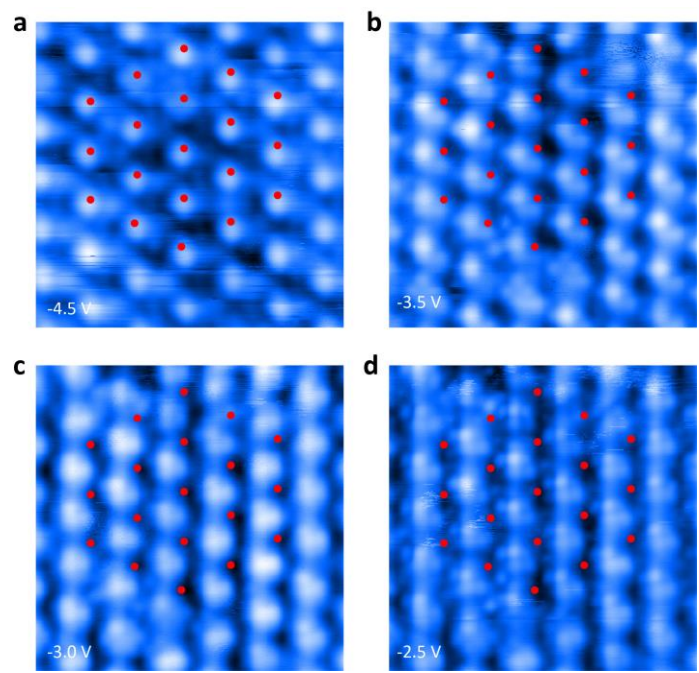

Figure S4. STM topographies (10 nm × 10 nm) of the Wigner crystal phase at various sample voltage as indicated. Analogous to the grid phase, the STM topographies rely strongly on the applied sample voltage and undergo a reversal of spatial contrast (c.f. Fig. S4a and Fig. S4d), suggestive of an electronic origin of the distorted hexagonal superstructure. The red dots marks the same positions to guide the eyes.

S5. Incommensurate stripe phase on LSCO films with varying thicknesses

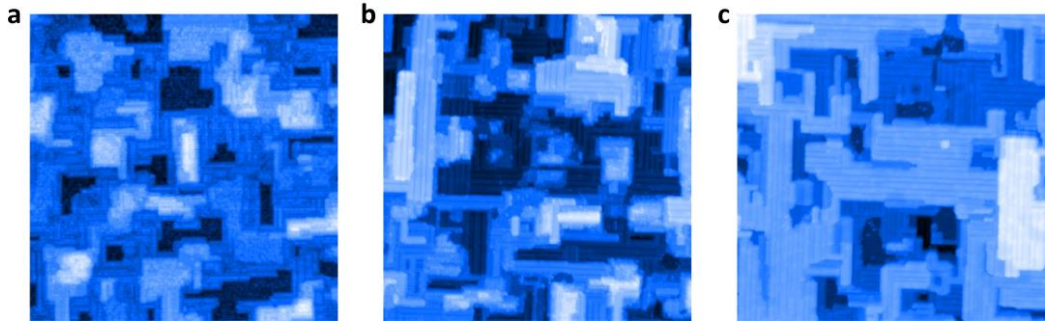

Figure S5. STM topographic images of incommensurate stripe phase on LSCO films with varying thicknesses. The consistent appearance of incommensurate stripe phase down to one UC hints at the minor role of epitaxial strain in the stabilization of charge orders in LSCO films. a, 1 UC, $V$ = -5 V, $I$ = 10 pA, 200 nm × 200 nm; b, 2 UC, $V$ = -5 V, $I$ = 10 pA, 200 nm × 200 nm; c, 3 UC, $V$ = -5 V, $I$ = 10 pA, 200 nm × 200 nm.

S6. Characterization of the LSCO films on LSAO substrate

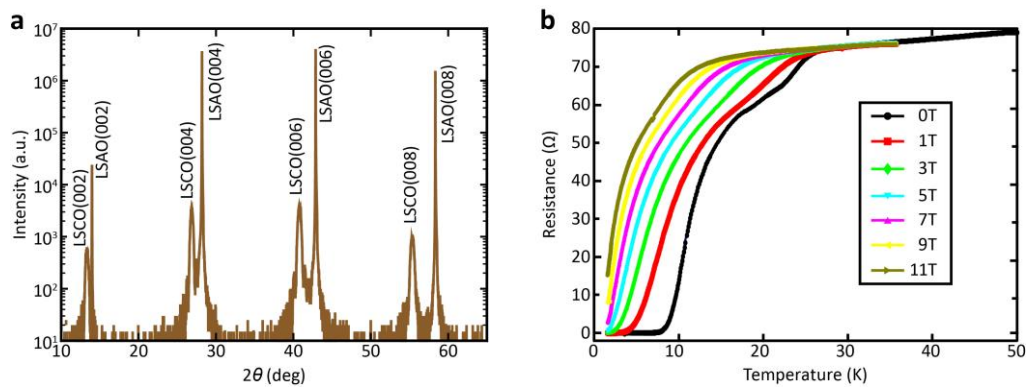

Figure S6. Characterization of epitaxial LSCO films on LSAO(001) substrate. a, XRD pattern of a representative LSCO epitaxial film grown on LSAO substrate. The Bragg peaks characteristic of both LSCO films and LSAO substrate are indicated. b, Resistance versus temperature curves under various magnetic fields perpendicular to the sample surface. The superconducting onset temperature $T_c$ under zero magnetic field is 26 K.

Using the identical growth recipe, we have also prepared Sr-doped LSCO films on insulating LSAO substrates. The subsequent X-ray diffraction measurements reveal similar crystal quality as those on $SrTiO_3$ substrates (Fig. S5a). However, the important distinction is that the epitaxial LSCO films prepared on LSAO substrates exhibit clear superconductivity (Fig. S5b), with a critical transition temperature $T_c$ of around 26 K. This hints at a competition scenario of charge orders and superconductivity in LSCO cuprates. Unfortunately, we failed to investigate the LSCO films on the insulating LSAO substrates by STM, because the superconducting LSCO films exhibit bad electrical contact with the sample holder, to which the STM bias voltage is applied. This merits a further experimental endeavor for characterization of LSCO/LSAO films in real space.